\RequirePackage{lineno}
\documentclass[aps,prd,twocolumn,showpacs,groupedaddress]{revtex4}
\usepackage{a4p,wrapfig,epsfig,epsf,graphics}
\usepackage{graphicx}
\usepackage{multirow}
\usepackage{mathrsfs}
\usepackage{amsmath}
\usepackage{atlasphysics}
\usepackage{mathptmx}
\usepackage{helvet}
\usepackage{multirow}
\usepackage{epstopdf}
\usepackage{subfigure}
\usepackage{verbatim}

\usepackage{hhline}

\newlength{\capindent}
\newlength{\capwidth}
\newlength{\figwidth}
\newcommand{\icaption}[2][!*!,!]{\hspace*{\capindent}%
  \begin{minipage}{\capwidth}
    \ifthenelse{\equal{#1}{!*!,!}}%
      {\caption{#2}}%
      {\caption[#1]{#2}}
      \vspace*{3mm}
  \end{minipage}}

\def \intlum {\int {\cal L} dt}
\newcommand{\Zgstar}{Z\kern -0.15em/\kern -0.15em\gamma^*}
\newcommand{\pp}{pp}
\newcommand{\invfb}{\rm fb^{-1}}
\newcommand{\Zg}{Z^0\kern -0.25em/\kern -0.15em\gamma^*}

\RequirePackage[colorlinks,breaklinks,pdftitle={ATLAS draft},pdfauthor={The ATLAS Collaboration}]{hyperref}  
\hypersetup{linkcolor=blue,citecolor=blue,filecolor=black,urlcolor=blue} 
\def\thistitle{Study of heavy-flavor quarks produced in association with top-quark pairs at $\sqrt{s}=7$~TeV using the ATLAS detector}
\def\thisabstract{Using a sample of dilepton top-quark pair ($\ttbar$)
  candidate events, a study is performed of the production of
  top-quark pairs together with heavy-flavor (HF) quarks, $\ttbar + b
  + X$ or $\ttbar + c + X$, collectively referred to as $\ttbar$ + HF. 
  The dataset used corresponds to an integrated luminosity of 4.7 $\invfb$
  of proton--proton collisions at a center-of-mass energy of $7~\TeV$
  recorded by the ATLAS detector at the CERN Large Hadron Collider. 
The presence of additional HF quarks in the $\ttbar$ sample is
inferred by looking for events with at least three $b$-tagged jets,
where two are attributed to the $b$-quarks from the $\ttbar$ decays
and the third to additional HF production. 
  The dominant background to $\ttbar$~+~HF in this sample is
  $\ttbar$~+~jet events in which a light-flavor jet is misidentified
  as a heavy-flavor jet. To determine the heavy- and light-flavor
  content of the additional $b$-tagged jets, a fit to the vertex mass
  distribution of $b$-tagged jets in the sample is performed. The
  result of the fit shows that 79 $\pm$ 14 (stat.) $\pm$ 22 (syst.) of
  the 105 selected extra $b$-tagged jets originate from HF quarks, three
  standard deviations away from the hypothesis of zero $\ttbar$~+~HF
  production.
  The result for extra HF production is quoted as
  a ratio ($R_{\rm HF}$) of the cross section for $\ttbar$~+~HF
  production to the cross section for $\ttbar$ production with at
  least one additional jet.  Both cross sections are measured in a
  fiducial kinematic region within the ATLAS acceptance.  $R_{\rm HF}$
  is measured to be 
[$6.2 \pm 1.1 (\text{stat.}) \pm 1.8 (\text{syst.})$]\% for jets with $\pt > 25~\gev$ and $|\eta| < 2.5$, in agreement with
the expectations from Monte Carlo generators.}

\begin{document}
\title{\thistitle}
\date{\today}

\begin{abstract}
\thisabstract
\end{abstract}

\maketitle

\section{Introduction}
\label{s:introduction}

In order to characterize the recently observed Higgs-like particle
($H$)~\cite{cmshiggs,atlashiggs}, quantities such as the Yukawa
coupling of the top quark and the Higgs boson need to be measured with
precision. For a Standard Model (SM) Higgs boson with a mass of
$125~\gev$, the decay mode with the largest branching ratio is $H
\rightarrow \bbbar$.  Thus, the channel with the largest yields for
studying $\ttbar + H$ production is $\ttbar + H$, $H \rightarrow
\bbbar$. Production of top-quark pair ($\ttbar$) events featuring
additional heavy-flavor (HF) $b$- and $c$-quarks, $\ttbar + b + X$ and
$\ttbar + c + X$, referred to as $\ttbar$ + HF, is the main
irreducible background to $\ttbar + H$, $H \rightarrow b\bar{b}$. A
study of $\ttbar$~+~HF production is useful to constrain models of
heavy-flavor quark production at the scale of the top-quark mass.
This analysis is also of interest because of the many potential
phenomena beyond the SM, such as composite Higgs
models~\cite{Montull:2012ik} and processes leading to final states
with four top
quarks~\cite{Degrande:2010kt,PhysRevD.51.3888,PhysRevD.78.074026,1126-6708-2008-04-087,1126-6708-2009-05-022,Guchait:2007jd},
that could produce additional heavy-flavor quarks in the $\ttbar$
candidate sample.

This paper describes a study of $\ttbar$ + HF production. Within the
SM, heavy-flavor quark pairs, $c\bar{c}$ and $b\bar{b}$, are expected
to be produced in association with $\ttbar$ mainly via gluon splitting
from initial- and final-state radiation~\cite{tt_hf_theory}. In
addition, the heavy-flavor content of the proton could lead to
$\ttbar$ final states with at least one additional HF quark, $\ttbar +
c$ and $\ttbar + b$. The data analyzed correspond to an integrated
luminosity of 4.7 $\ifb$ at a center-of-mass energy of
$\sqrt{s}~=~7~\tev$ produced at the Large Hadron Collider (LHC) and
recorded in 2011 with the ATLAS detector.

This analysis is performed on $\ttbar$ dilepton candidate events in
which each top quark decays to a $b$-quark and a $W$ boson, which
subsequently decays to a neutrino and an isolated, charged lepton.
The dilepton signature is selected for this measurement because it is
relatively background free and precludes a third $b$-tagged jet from a
hadronically decaying $W$ boson, predominantly via $W \rightarrow
s\bar{c}$. The $\ttbar$~+~HF signal region is the subset of these
events with three or more jets identified as containing HF quarks
($b$-tagged jets, or $b$-tags). However, jets without HF quarks may
also be $b$-tagged, so that care must be taken to properly identify
the flavor composition of the $b$-tagged jets in the sample. Two
$b$-tagged jets from each event are presumed to originate from the
$b$-quarks from top-quark decays, $\ttbar\rightarrow W^+ b W^-
\bar{b}$. Therefore, all events in the signal region have at least one
additional $b$-tag either from a $b$- or $c$-quark jet, or from a
light-quark or gluon jet that was misidentified. The latter are
referred to as light-flavor or LF jets.

Due to limited data statistics and discrimination between $b$- and
$c$-jets, the sum of $b$-quark and $c$-quark jet rates is measured. 
Information about the composition of $\ttbar + b + X$ and $\ttbar + c
+ X$ in $\ttbar + $~HF is nevertheless required for the total
correction due to acceptance, which is different for $b$- and
$c$-quark jets. 
The composition is estimated with Monte Carlo
simulation and tested in the data.

From the measurement of the fraction of jets with heavy flavor
content, the cross section for $\ttbar$ production with
at least one additional HF jet can be extracted.  
To reduce some systematic
uncertainties, the result is quoted as a ratio, termed $R_{\rm HF}$,
of the cross section for $\ttbar$ production with at least one
additional HF jet to the cross section for $\ttbar$ production with at
least one additional jet ($\ttbar + j$), regardless of flavor. The
measurement of $\ttbar + j$ production is performed in dilepton
$\ttbar$ candidate events with at least three jets, at least two
of which are $b$-tagged and assumed to come from top-quark decays.

The paper is organized as follows. The ATLAS detector is briefly
described in Sec.~\ref{s:detector}. The data and Monte Carlo samples
used in the analysis are described in Sec.~\ref{s:mc}, followed by a
description of the event selection in Sec.~\ref{s:selection}.  The
definition of the fiducial phase space used in the measurement of
$R_{\rm HF}$, and the calculation of acceptances and
efficiencies are presented in Sec.~\ref{s:fiducial}. In
Sec.~\ref{s:expectation}, observed and expected numbers of events with
$\ge~3$ $b$-tagged jets are shown. Section~\ref{s:template} describes
a fit to the vertex mass distribution of $b$-tagged jets in these data
events to extract the fraction of HF jets produced in association with
$\ttbar$. A discussion of the systematic uncertainties of the
measurement is presented in
Sec.~\ref{s:systematics}. Section~\ref{s:results} shows the result of
the measurements, followed by conclusions in Sec.~\ref{s:conclusions}.

\section{The ATLAS detector}
\label{s:detector}

A detailed description of the ATLAS detector can be found
elsewhere~\cite{Aad:2008zzm}. The innermost part of the detector is a
tracking system that is immersed in a 2 T axial magnetic field and
that measures the momentum of charged particles.  The inner detector
comprises a silicon pixel detector, a silicon microstrip detector, and
a transition radiation tracker, providing tracking capability within
the pseudorapidity range ${|\eta|<2.5}$~\cite{atlas_coordinates}.  The
tracking system is also used to identify the displaced secondary
vertex that is formed by hadrons containing a $b$- or $c$-quark.
Calorimeter systems, which measure the electron, photon, and hadron
energies, reside outside the inner detector and cover the region
$|\eta|<4.9$.  Outside the calorimeters there is a muon spectrometer
that is used to identify and measure the momentum of muons in an
azimuthal magnetic field in the region $|\eta|<2.7$.  To reduce the
data rate, a three-level trigger system selects the potentially
interesting events that are recorded for offline analysis.

\section{Data and Monte Carlo samples}
\label{s:mc}

The total integrated luminosity for the analyzed data sample is 4.7
$\ifb$ at a center-of-mass energy of $\sqrt{s}~=~7~\tev$. During the
2011 data-taking period the instantaneous luminosity of the LHC
increased, causing the average number of simultaneous inelastic $\pp$
interactions per beam crossing (pile-up) at the beginning of a $\pp$
fill to increase from about 6 to 17.  Multiple $pp$ interactions can
occur either in the same bunch crossing as the primary vertex (termed
`in-time pile-up') or in an adjacent bunch crossing (termed
`out-of-time pile-up'). To account for these effects, all Monte Carlo
simulated events are overlaid with additional inelastic events
generated with the {\sc Pythia} AMBT1 tune~\cite{Aad:2010ac}, and the
distribution of the number of vertices in the simulation is
reweighted to match the distribution of the number of additional
interactions per bunch crossing measured in the data.

Monte Carlo simulation is used to study signal and background
processes. 
Inclusive $\ttbar$ production and dedicated $\ttbar$ + HF samples are simulated using the multi-leg
matrix-element 
generator {\sc Alpgen} v2.13~\cite{MAN-0301} with the CTEQ6L1~\cite{Stump:2003yu}
parton distribution function (PDF) set.  Parton showering and
hadronization are performed by {\sc Herwig} v6.520~\cite{COR-0001}.
Effects due to the mass of the heavy-flavor quarks are included by
default in {\sc Alpgen}. 
In these samples, additional jets (including heavy-flavor) can also be
produced in the parton shower. 
The MLM~\cite{MAN-0301} parton--jet matching scheme is applied to
avoid double counting of configurations generated by both the parton
shower and the leading-order (LO) matrix-element (ME) calculation. In
addition, overlap between $\ttbar$ events with HF quarks that
originate from ME production and those that originate from the parton
shower is removed.  This heavy flavor overlap removal (HFOR) is based
on the $\Delta R_{qq}$~\cite{deltar} between simulated HF quarks. The
event is taken from the ME calculation if it contains two well
separated HF quarks ($\Delta R_{qq}>0.4$). The event is taken from the
parton shower calculation if it contains two collinear HF quarks
($\Delta R_{qq}<0.4$).

To study the effect of different fixed-order calculations and matching
schemes, samples of top-quark pair events are also generated using
{\sc Powheg}~v1.01 and showered with {\sc Herwig}.  In this sample the
$\ttbar$ process is described at next-to-leading order (NLO), while
the extra jets are described at LO. For each sample showered with {\sc
  Herwig}, {\sc Jimmy} v4.31~\cite{JButterworth:1996zw} and the AUET1
tune~\cite{PUB-2010-014} are used to simulate the underlying event and
to model various soft interactions. To assess the effect of different
parton shower models, a sample is generated using {\sc Alpgen} v2.14
with the {\sc Pythia} v6.425~\cite{Sjostrand:2006za} parton shower and
hadronization, using the CTEQ5L PDF set~\cite{CTEQ5}. The uncertainty
associated with the CTEQ6L1 PDF set is evaluated with an envelope
calculated using the uncertainty set from the NLO PDF
MTSW2008nlo68cl~\cite{Martin:2009iq}, and an additional term to
account for the difference between the central values of the LO and
NLO calculations.

Initial- and final-state radiation (ISR/FSR) variations are studied
using samples generated with {\sc AcerMC} v2.0~\cite{Kersevan:2013ji}
interfaced with {\sc Pythia} v6.2. In these samples the parameters
that control the amount of ISR/FSR are set to points consistent with
the PERUGIA Hard/Soft tune~\cite{Skands:2010ak} in a range constrained
by current experimental data~\cite{ATLAS:2012al}.

In all samples the top-quark mass is set to $m_t = 172.5~\gev$.  
The cross section for Standard Model $\ttbar$ production at this mass is
calculated using the approximate next-to-next-to-leading-order (NNLO)
QCD calculation described in~\cite{HATHOR}.
 
Background samples from the production of $W$ and $Z$ bosons are
generated using the CTEQ6L1 PDFs with {\sc Alpgen}, which is
interfaced to {\sc Herwig}
for parton showering and hadronization; the {\sc Alpgen} matrix
elements include diagrams with up to five additional partons. 
Separate samples of $W+b\bar{b}$ and $Z+b\bar{b}$ events are generated. 
The overlap between jets from the parton-shower and the matrix-element in the
$n$ and $n+1$ jet multiplicity samples is removed for the $W$+jets and
$Z$+jets samples in the same manner as for the $\ttbar$ samples. 
Single top-quark production is modeled using {\sc AcerMC} in the $t$-channel and {\sc MC@NLO} v3.41~\cite{Frixione:2002ik} for the $Wt$- and $s$-channels. 
Diboson ($WW$, $WZ$, and $ZZ$) production is modeled using {\sc Alpgen} interfaced with {\sc Herwig}. 
Less than 0.5\% of the expected yield in the $\ttbar$ + HF sample comes from the associated production of $\ttbar + W/Z$ and $\ttbar + H$, and these processes are thus neglected in this analysis.

The resulting generated samples are passed through a {\sc Geant4}
simulation~\cite{AGO-0301} of the ATLAS detector~\cite{:2010wqa}.
Events are then reconstructed in the same manner as the data.

\section{Event selection}
\label{s:selection}

Events for the analysis are selected by at least one of the
high-$\pt$~\cite{atlas_coordinates} single-electron or single-muon triggers, 
as described in Refs.~\cite{ATLASTrigPerformance} and~\cite{ATLAS:2012yna}. 
The single-electron triggers are
based on calorimeter energy deposits, shower shape, and matching track quality
constraints, while the single-muon triggers are based on a
reconstructed track in the muon spectrometer that matches a track
found in the inner detector. To ensure a final trigger rate that is
compatible with the ATLAS data acquisition system, a minimum $\pt$
threshold for the electron and muon triggers is used. The $\pt$
threshold for the muon trigger is $18~\gev$. For the electron trigger,
the threshold is $20~\gev$ or $22~\gev$, depending on the data-taking
period due to varying LHC luminosity conditions.

The selected events are required to contain a reconstructed primary
vertex with at least five associated tracks with $\pt>0.4~\GeV$. Event
reconstruction makes use of electrons ($e$), muons ($\mu$), jets, and
missing transverse momentum ($\met$). Electrons are reconstructed by
matching energy deposits in the electromagnetic calorimeter with
tracks in the inner detector, and are required to have $\pt > 25~\gev$
and $|\eta| < 2.47$, excluding the transition region between the
barrel and endcap calorimeters at
1.37$<|\eta|<$1.52~\cite{ATLASElectronPerformance}.  Muons are
reconstructed by matching tracks in the inner detector with tracks
measured in the muon spectrometer, and are required to have $\pt >
20~\gev$ and $|\eta| < 2.5$.

Tight isolation cuts are applied
to both the electron and muon candidates to reduce the number of
identified leptons ($e$, $\mu$) that come from non-prompt (non-$W/Z$)
sources and from misidentified hadrons. 
For electrons, the $\et$ deposited in the calorimeter cells in a cone
in $\eta$-$\phi$ space of radius $\Delta R=0.2$ around
the electron position is summed, and the $\et$ due to the electron
is subtracted. The scalar sum of track transverse momenta in a cone of
$\Delta R=0.3$, excluding the electron,
 is also measured. 
Cuts parametrized by the electron $\eta$ and $\et$ are made on these
two isolation variables to ensure a constant efficiency over the
entire ($\eta$, $\et$) range. 
For
muons, the corresponding calorimeter isolation energy in a cone of
$\Delta R=0.2$ is required to be less than $4~\gev$, and the scalar
sum of track transverse momenta in a cone of $\Delta R=0.3$ is
required to be less than $2.5~\gev$ after subtraction of the muon
$\pt$.

\begin{figure*}[t]
\centering
\subfigure[Zero $b$-tagged jets]{\includegraphics[scale=0.25]{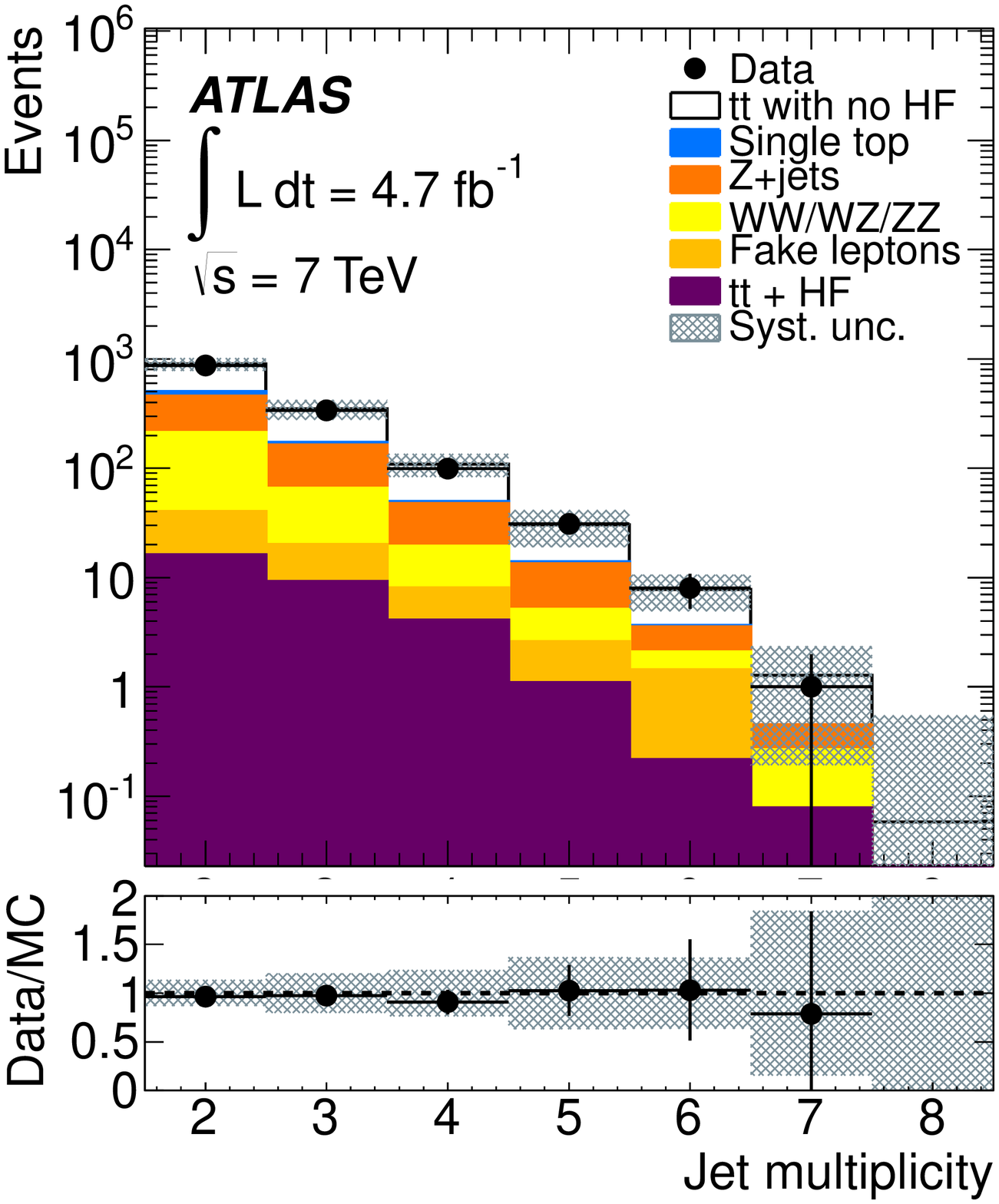}}
\subfigure[One $b$-tagged jet ]{\includegraphics[scale=0.25]{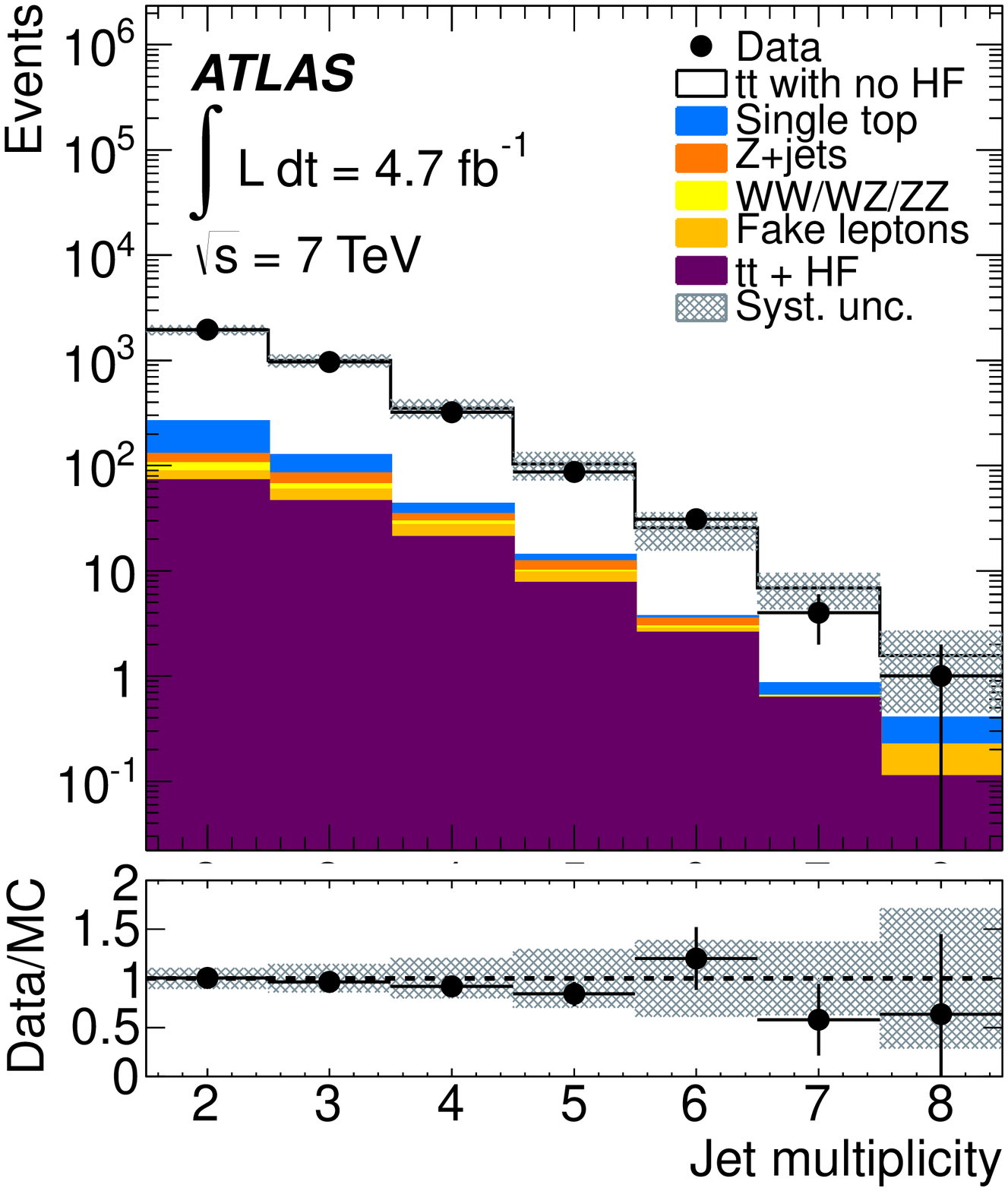}}
\subfigure[Two $b$-tagged jets]{\includegraphics[scale=0.25]{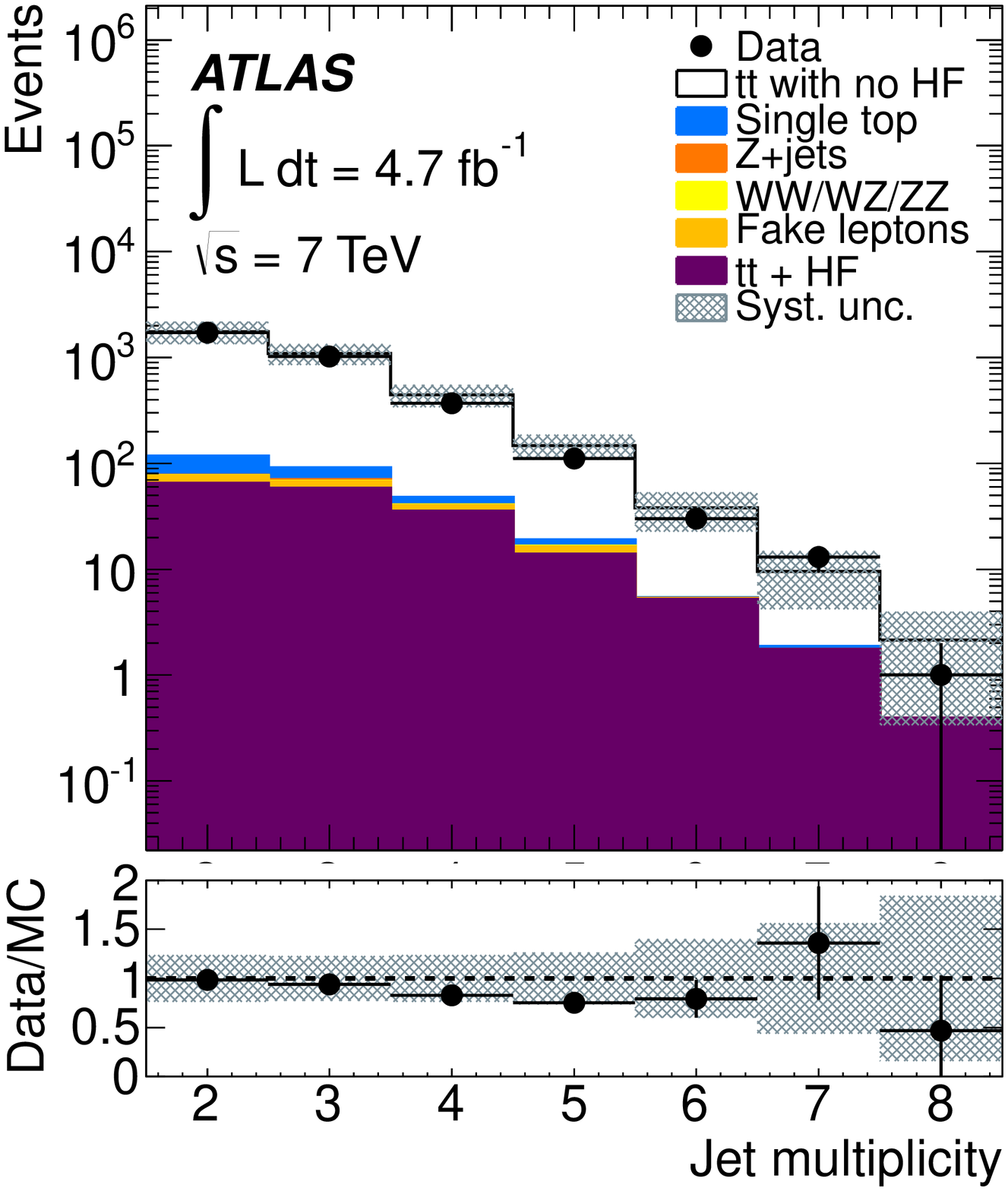}}
\caption[Selected jet multiplicity distributions in the $\ttbar$
  control region] {Jet multiplicity distributions in dilepton $\ttbar$
  candidate events with (a) zero, (b) one, or (c) two $b$-tagged jets
  for the sum of $ee$, $\mu\mu$ and $e\mu$ channels. The lower plots
  show the ratio between the data and the Monte Carlo predictions in
  each bin. Uncertainties are statistical and systematic. The last bin
  contains any overflow.}
\label{f:njets}
\end{figure*}

Jets are reconstructed from clustered energy deposits in the
calorimeters with the anti-$k_t$~\cite{antikt} algorithm with a radius
parameter $R\,=\,0.4$~\cite{Aad:2011he}.  Jets selected for the
analysis are required to have $\pt > 25~\gev$ and $|\eta| < 2.5$. In
order to reduce the background from jets originating from pile-up
interactions, additional selection criteria are applied to the
fraction of the jet's $\pt$ ($JVF$) carried by tracks originating from
the primary vertex, $JVF > 0.75$.

The transverse momentum of neutrinos produced in the top-quark decays,
measured as $\met$, is inferred by balancing the vector sum of all
visible transverse momenta. Specifically, the $\met$ is constructed
from the vector sum of all calorimeter cell energies contained in
topological clusters~\cite{Aad:2011he} with $|\eta|<4.5$, projected
onto the transverse plane.  Contributions to the $\met$ from the
calorimeter cells associated with jets are taken at the corrected
energy scale that is used for jets, while the contribution from cells
associated with electrons is substituted by the calibrated transverse
momentum of the electron.  The contribution to the $\met$ from the
$\pt$ of muons passing the selection requirements is also included.

The $b$-tagging algorithm~\cite{ATLAS-CONF-2012-040,ATLAS-CONF-2012-043} employed for
this analysis uses impact parameter and vertex position measurements
from tracks in the inner detector as inputs to a neural network.  The
$b$-tagging efficiency was calibrated in a multi-jet data sample where
at least one jet contains a muon~\cite{ATLAS-CONF-2012-043}. The
$c$-tagging efficiency was calibrated in a data sample with
reconstructed $D^\ast$ mesons~\cite{ATLAS:2012mma}.  For this
analysis, $b$-tagged jets are required to satisfy a selection that is
75\% efficient for $b$-quark jets, approximately 30\% efficient for
$c$-quark jets, and rejects light-flavor jets by a factor of
approximately 35 in simulated $\ttbar$ events.  
In this paper, a `$b$-tag' (or a `$b$-tagged jet') refers to any jet
passing this selection, regardless of flavor. A `$b$-jet', by
contrast, refers to a jet (which may or may not be $b$-tagged) which
contains a $b$-quark. Similarly, `$c$-jet' and `HF jet' are statements
of the flavor composition of the jet, not whether the jet is
$b$-tagged. 
Three distinct subsets of the selected $b$-tagged jets with different 
$b$-jet purity are used in the measurement of $\sigma_{\rm
  fid}(\ttbar$ + HF), as described in Sec.~\ref{s:template}.

Dilepton $\ttbar$ candidate events are selected by requiring exactly two
opposite-sign leptons and at least two jets. To reduce the background
from $\Zgstar$ processes, events with like-flavor leptons are required
to have $\met$ above $60~\gev$ and a dilepton invariant mass satisfying
$|m_{\ell^+\ell^-} - m_Z| > 10~\GeV$.  For events with one electron
and one muon, the scalar sum of the lepton and jet transverse momenta
is required to be above $130~\gev$ to reduce the backgrounds from
$\Zgstar \rightarrow \tau^+\tau^-$, as well as $WW$, $WZ$, and $ZZ$
processes.  This set of selection criteria is termed the `nominal'
$\ttbar$ selection criteria.  The measurement of $\ttbar$ + HF
production is carried out in the subset of these events that contain
three or more $b$-tagged jets, whereas the measurement of $\ttbar$
production with at least one additional jet is performed in the subset
with at least three jets, at least two of which are $b$-tagged.

Using the nominal selection criteria described above, data and Monte
Carlo events are compared in three control regions: dilepton $\ttbar$
candidate events with zero, one, or two $b$-tagged jets.
Data-to-simulation normalization corrections 
are applied to Monte Carlo simulation samples when calculating
acceptances to account for observed differences in predicted and
observed trigger and lepton reconstruction efficiencies, jet flavor
tagging efficiencies and mistag rates, as well as jet and lepton
energy scales and resolutions.  In Fig.~\ref{f:njets}, the jet
multiplicity distributions in the three regions are compared to Monte
Carlo predictions. Agreement is observed within uncertainties.

\section{Definition of the fiducial phase space and calculation of correction factors}
\label{s:fiducial}

To allow comparison of the analysis results to theoretical
predictions, the measurement is made within a fiducial phase
space. The fiducial volume is defined in Monte Carlo simulation by
requiring two leptons ($e$, $\mu$) from the $t \rightarrow Wb
\rightarrow \ell \nu b$ decays (including electrons and muons coming
from $\tau \rightarrow \ell\nu\nu_\tau$) with $\pt > 25~ (20)~\gev$
for $e$ ($\mu$), and $|\eta| < 2.5$ as well as three or more jets with
$\pt > 25~\gev$ and $|\eta| < 2.5$.

In the simulation, jets are formed by considering all particles with a
lifetime longer than 10 ps, excluding muons and neutrinos. Particles
arising from pile-up interactions are not considered.  For the
determination of the $\ttbar$ + HF fiducial cross section,
$\sigma_{\rm fid}(\ttbar + {\rm HF})$, three or more jets are required
to match a $b$- or $c$-quark, two of which must match a $b$-quark from
top-quark decay. All simulated $b$- and $c$-quarks that were generated
with $\pt > 5~\gev$ are considered for the matching, and are required
to satisfy $\Delta R$(quark,~jet)~$<$~0.25. Jets that match both a
$b$- and a $c$-quark are considered as $b$-jets. 
For the calculation of $\sigma_{\rm fid}(\ttbar+j)$ three or more jets
are required, two of which must contain a $b$-quark from top-quark
decay.

Each fiducial cross section is determined using measured quantities
from the data, and a correction factor derived from the Monte Carlo
simulation.  The ratio of cross sections is defined as:

\begin{equation}
R_{\rm HF} = \frac{\sigma_{\rm fid}(\ttbar + {\rm HF})}{\sigma_{\rm fid}(\ttbar+j)} \nonumber
\end{equation}

\noindent The fiducial cross section for $\ttbar$ + HF production is determined from:

\begin{equation}
\sigma_{\rm fid}(\ttbar + {\rm HF}) = \frac{N_{\rm HF}}{\intlum \cdot \epsilon_{\rm HF}};
\label{eq:xsechf}
\end{equation}

\noindent where $N_{\rm HF}$ is the number, after background
subtraction, of $b$-tags from HF jets observed in the data, in
addition to the two $b$-jets from top-quark decays. The integrated
luminosity of the sample is denoted as $\intlum$, and $\epsilon_{\rm
  HF}$ is a correction factor taken from Monte Carlo simulation that
converts the number of observed $b$-tags from additional HF jets to
the number of events in the signal fiducial volume. This correction
factor includes the acceptance within the fiducial region, the reconstruction efficiency, and a
factor to account for the multiplicity of extra $b$-tagged HF jets per
$\ttbar$+HF event in the signal region. 
This correction factor is different for $\ttbar+b+X$
($\epsilon_b$) and $\ttbar+c+X$ ($\epsilon_c$), and thus
$\epsilon_{HF}$ is determined as a weighted sum of these two
contributions. 
The weight used to form the sum is the fraction of $\ttbar + \rm{HF}$
events in the fiducial volume which contain additional $b$-jets as
opposed to $c$-jets. 
This fraction is termed $F_{b/{\rm HF}}$. 
The total correction factor ($\epsilon_{\rm HF}$) is
calculated as:
\begin{equation}
\epsilon_{\rm HF} = F_{b/{\rm HF}} \cdot \epsilon_b + (1 - F_{b/{\rm HF}}) \cdot \epsilon_c \nonumber
\end{equation}
\noindent The denominator for $R_{\rm HF}$, $\sigma_{\rm fid}(\ttbar+j)$, is
computed using a similar prescription:
\begin{equation}
\sigma_{\rm fid}(\ttbar + j) = \frac{N_j}{\intlum \cdot \epsilon_j};
\label{eq:xsec}
\end{equation}
\noindent where $N_j$ is the yield of dilepton events in data
with at least three jets, at least two of which are $b$-tagged, 
and
$\epsilon_j$ is the $\ttbar + j$ acceptance factor calculated from the
Monte Carlo simulation.  The acceptance calculation for each fiducial
cross section assumes that all $b$-tagged jets are from real HF
quarks. Events with $b$-tagged jets from LF quarks are treated as a
background, and subtracted when computing both $N_{\rm HF}$ and $N_j$.

The {\sc Alpgen} + {\sc Herwig} Monte Carlo sample 
predicts $\epsilon_b$ = 0.19, $\epsilon_c$ = 0.06, and $F_{b/{\rm
    HF}}$ = 0.31. The total correction factor is thus predicted to be
$\epsilon_{\rm HF}$ = 0.106 $\pm$ 0.005 (stat.)  for $\sigma_{\rm
  fid}(\ttbar + {\rm HF})$. For $\sigma_{\rm fid}(\ttbar +j)$ the
acceptance factor is calculated to be $\epsilon_j$ = 0.129 $\pm$ 0.001
(stat.).

The prediction for $R_{\rm HF}$ from the {\sc Alpgen} + {\sc Herwig}
Monte Carlo sample is 3.4\%. 
The value obtained from the {\sc Powheg}~v1.01~\cite{Alioli:2010xd}
generator showered with {\sc Herwig}~\cite{COR-0001} 
is $R_{\rm HF}$ = 5.2\%, with $F_{b/{\rm
    HF}}$ = 0.34. While this $R_{\rm HF}$ value is different to that from
{\sc Alpgen} + {\sc Herwig}, the predicted $F_{b/{\rm HF}}$ values are similar.
Furthermore, a parton-level study using {\sc MadGraph5}
v1.47~\cite{Maltoni:2002qb} gives $F_{b/{\rm HF}} = 0.29$. 
The value of $F_{b/{\rm HF}}$ is also stable when different showering algorithms are used:  
the {\sc Alpgen} + {\sc Pythia} Monte Carlo sample predicts 
a value of $F_{b/{\rm HF}}$ = 0.32, in good agreement with the
prediction when {\sc Herwig} is used. 
Based on comparison of these predictions for $F_{b/{\rm HF}}$, a symmetric 10\%
Monte Carlo systematic uncertainty is assigned, $F_{b/{\rm HF}} = 0.31
\pm 0.03$. 
The prediction of $F_{b/{\rm HF}}$ is also tested in data (see
Sec.~\ref{s:results}).

\section{Expected signal and background yields}
\label{s:expectation}

Table~\ref{os-3+-table} shows the number of events with $\ge~3$
$b$-tagged jets expected in the Monte Carlo simulation from dilepton
$\ttbar$ production and from various background sources. At this
point, no distinction is made between events with a true additional
HF jet and those containing a mistagged LF jet.  The number of
observed events is also shown.  While Monte Carlo simulation is used
to estimate $\ttbar$ + HF event rates and kinematic features,
data-driven methods and Monte Carlo simulation are both used to
estimate background processes, as detailed below. 

Background processes containing real $b$-jets and leptons, such as
single top-quark, $\Zgstar$ + jets, and diboson ($WW$, $WZ$, and $ZZ$)
production, are estimated using Monte Carlo simulation. Contributions
from diboson production are found to be negligible.

A major source of background comes from $\ttbar$ events in which one
or more of the $b$-tagged jets is from a mistagged LF jet.  This
background is estimated using Monte Carlo simulation for the
measurement of $\sigma_{\rm fid}(\ttbar+j)$. However, in the
measurement of $\sigma_{\rm fid}(\ttbar$ + HF), the final $\ttbar$ +
LF background is determined by a fit to the vertex mass distribution
of $b$-tagged jets in data, as explained in Sec.~\ref{s:template}.

Background from events in which at least one of the leptons is either
non-prompt (originating from e.g. a photon conversion or $b$-quark
decay) or is a misidentified hadron, is estimated using data and Monte
Carlo simulation. For instance, $W$ + jets, multi-jet, and $\ttbar$
events with one hadronically decaying $W$ boson can contribute in this
way. This contribution is determined by scaling the yield of events in
the data with a pair of same-sign leptons by the ratio of
opposite-sign to same-sign yields ($R_{\rm OS/SS}$) obtained in Monte
Carlo simulation.  The opposite-sign to same-sign ratio is determined
separately for the three dilepton channels, and found to be 1.3 $\pm$
0.1 (stat.)  $\substack{+1.8\\-1.3}$ (syst.) for $e^+e^-$ events,
1.2 $\pm$ 0.1 (stat.) $\pm$ 0.7 (syst.) for $\mu^+\mu^-$ events,
and 1.2 $\pm$ 0.1 (stat.) $\pm$ 0.5 (syst.)  for events with one
electron and one muon. The systematic uncertainty takes into account
the unknown relative mixture of fake-lepton sources (photon
conversions, $b$- and $c$-hadron decays, or misidentified hadrons) in
the $R_{\rm OS/SS}$ calculation. 
Since the central value of the prediction for this background is zero events,
only variations in $R_{\rm OS/SS}$ that lead to larger background
predictions are considered in the systematic uncertainty calculation.
This method for estimating the background due to events with fake
leptons is validated in a control sample of dilepton events with less
restrictive lepton identification requirements and no isolation criteria.

\begin{table}[!tb]
  \caption{Observed and expected number of events in the signal region
    (i.e. with $\geq$ 3 $b$-tagged jets).  Uncertainties on
    individual components are statistical only.  For the total
    expectation, systematic uncertainties are included.  
  } 
\begin{tabular}{c|c}
\hline
\hline
Process & Number of events \\
\hline
$\ttbar$ & 106.7 $\pm$ 3.4 \\
Single top & 2.2 $\pm$ 0.5 \\
$Z$ + jets & 0.2 $\pm$ 0.1 \\
Fake leptons & 0 $\substack{+5\\-0}$\\
Total expectation & 109 $\substack{+6\\-3}$ (stat.) $\pm$ 35 (syst.)\\
\hline
Data & 106\\
\hline
\hline
\end{tabular}
\label{os-3+-table}
\end{table}

The dominant uncertainties on the total yield in
Table~\ref{os-3+-table} come from the jet energy scale, $b$-tagging
efficiency, parton showering model, and initial- and final-state
radiation.

\section{Template Fit}
\label{s:template}

For the measurement of $\sigma_{\rm fid}(\ttbar$ + HF), the fraction
of heavy-flavor jets produced in association with $\ttbar$ is
extracted by performing a binned maximum-likelihood fit on the
displaced-vertex mass distribution using all $b$-tagged jets in the
events with $\ge 3$ $b$-tagged jets. Although the final result is for
both flavors combined, the fit includes separate $b$- and $c$-quark
components to improve the determination of the LF fraction, and to
test the Monte Carlo prediction for $F_{b/{\rm HF}}$, which is used
for the calculation of the correction factor described in
Sec.~\ref{s:fiducial}.  This displaced-vertex mass is constructed from
the inner detector tracks associated with the secondary vertex using
the algorithm described in Ref.~\cite{ATLAS:2010psx}. 
While the presence of a displaced vertex is an indication that a jet
contains a $b$-quark, a jet may be $b$-tagged even if no vertex is
reconstructed.  In this case, the vertex mass is undefined. These jets
are assigned a mass value of `$-1~\GeV$' and they are included in the
fit to the displaced-vertex mass distribution. Keeping the events
without a reconstructed vertex improves the discrimination between
heavy-flavor and light-flavor jets.

While the vertex mass is a powerful discriminant,
Monte Carlo studies indicate that the sensitivity on the fitted
fraction of LF jets increases when the jet $\pt$ is used as an additional
discriminant. Considering only the statistical uncertainty,
it is seen that a fit with both jet $\pt$ and vertex mass is approximately half
a standard deviation more sensitive than a fit with only the vertex mass.
It was thus decided to define a two-dimensional probability density
function, termed a `template,' for the fit using the vertex mass and jet $\pt$.  
\begin{table}[!ht]
\caption{Summary of the $b$-tagging efficiencies for $b$-jets, $c$-jets, and
  light-flavor jets for the three {\it mutually exclusive} $b$-tagging
  selections used in the vertex mass template fit.}
\centering
\footnotesize
\begin{tabular}{c|c|c|c}
\hline
\hline
$b$-purity & $b$-jet efficiency & $c$-jet efficiency & light-flavor efficiency\\
\hline
High   & 60\% & 17\% & 0.43\% \\
Medium & 10\% &  7\% & 1.00\% \\
Low    &  5\% &  6\% & 1.33\% \\
\hline
\hline
\end{tabular}
\label{t:tagging_efficiency}
\end{table}

The fit is performed simultaneously in three {\it mutually exclusive}
bins of $b$-jet purity, defined by different ranges of the $b$-tagging
neural network output value. Certain values of the neural network
output, termed `operating points', are defined by the average $b$-jet
selection efficiency resulting from the applied selection.  In this
analysis, operating points of 60\%, 70\% and 75\% efficiency are used
to define the boundaries of the $b$-jet purity bins.

The first bin uses only the tightest calibrated operating point
(60\%), contains the highest-purity sample of $b$-jets (referred to as
`high purity'), and has a $b$-tagging efficiency of 60\% for
$b$-jets. The second bin (referred to as `medium purity') requires a
$b$-tag selection between the tightest and second tightest (70\%)
operating points, and contains a larger fraction of LF jets and
$c$-jets. The efficiency for this bin is 10\% for $b$-jets, i.e. the
difference between the 70\% and 60\% operating points. The final bin
(`low purity') requires a $b$-tag selection between the second (70\%)
and third operating point (75\%), and contains the largest fraction of
LF jets.  The efficiency for this bin is 5\% for $b$-jets.
The $b$-tagging
efficiencies for $b$-jets, $c$-jets, and light-flavor jets for each
selection are given in Table~\ref{t:tagging_efficiency}.

All three classes of $b$-tag purity are used in the analysis so that a
jet is considered `$b$-tagged' if it satisfies any of these criteria.
The discrimination power between LF and $c$-jets is greatly improved by
using three (as opposed to one) classes of $b$-purity. 
The vertex mass
distributions for all $b$-tagged jets in events passing the nominal
$\ttbar$ selection criteria are shown in Fig.~\ref{f:vertex_mass} to
confirm that (a) the data are well described by the Monte Carlo
simulation, (b) and the $b$-jet, $c$-jet and LF-jet fractions are
different in the three purity selections.  For the purpose of illustration, 
the normalization of the $b$-jet, $c$-jet, and LF-jet components is taken from 
Monte Carlo simulation.

\begin{figure*}[htbp]
\centering
\subfigure[~High-purity $b$-tagged jets  ]{\includegraphics[scale=0.34]{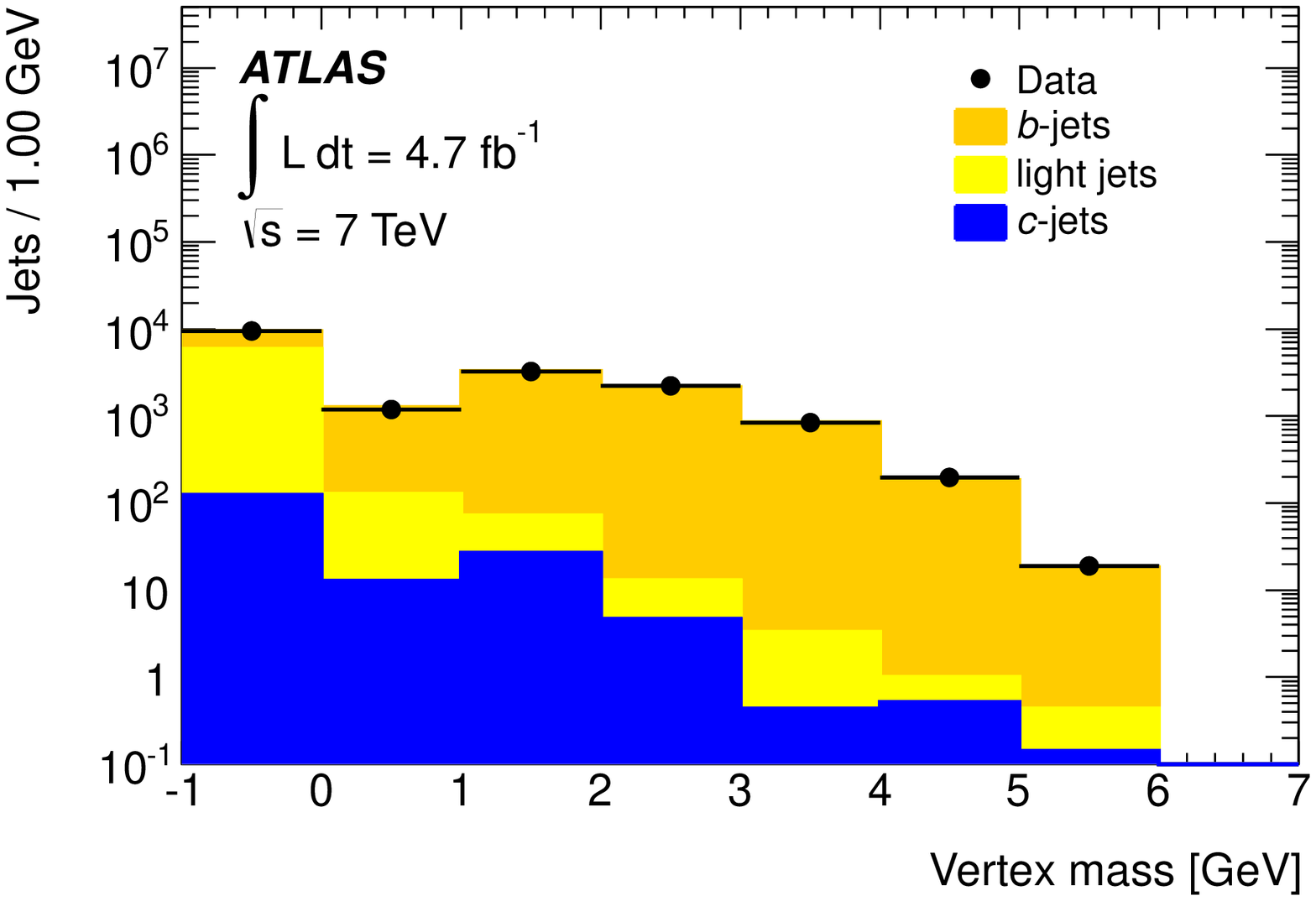}}
\subfigure[~Medium-purity $b$-tagged jets]{\includegraphics[scale=0.34]{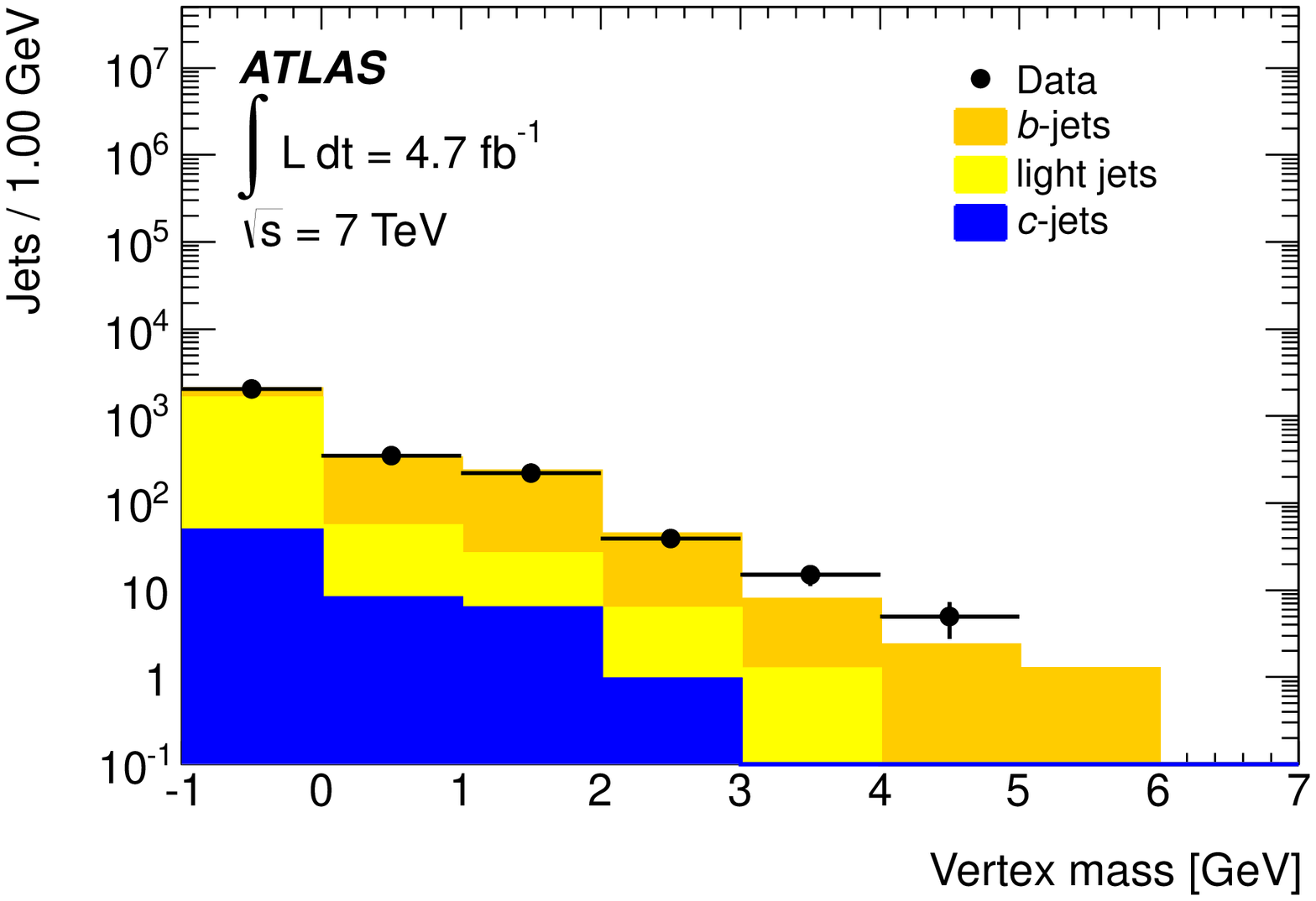}}
\subfigure[~Low-purity $b$-tagged jets   ]{\includegraphics[scale=0.34]{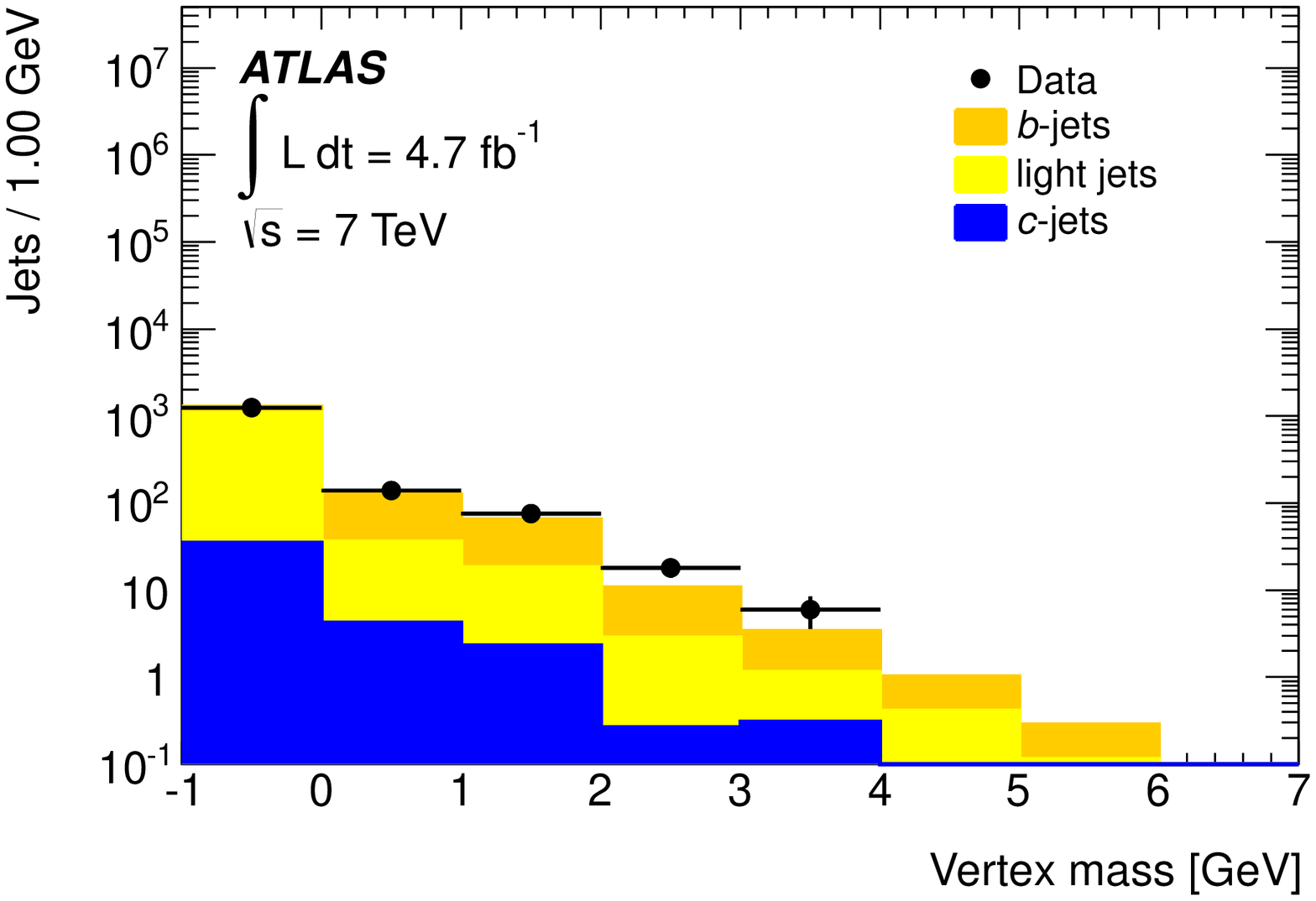}}
\caption[Vertex mass distribution] {Vertex mass distributions for
  all $b$-tagged jets in data events satisfying the nominal dilepton
  $\ttbar$ event selection, with no requirement on $b$-tagged jet
  multiplicity, compared to Monte Carlo predictions. By
  definition, jets with no reconstructed secondary decay vertex are
  assigned to the `$-1~\GeV$' bin. 
}
\label{f:vertex_mass}
\end{figure*}

The template fit has five components: $b$-jets from top-quark decays,
non-$\ttbar$ background, extra $b$-tagged jets from $b$-quarks, extra
$b$-tagged jets from $c$-quarks, and light-flavor $b$-tagged jets. The
template for $b$-jets from top-quark decays is obtained from the data
in $\ttbar$ dilepton events with exactly two $b$-tags.  Monte Carlo
simulation indicates that 97\% of $b$-tagged jets in $\ttbar$ dilepton
events with exactly two $b$-tags come from the decay of the top quark.
To account for this in the shape of the data template, a template for
$b$-tags not from the top-quark decays is derived from the $\ttbar$
Monte Carlo simulation, and subtracted with a 3\% relative
normalization from the data template. 
In the fit, the normalization for the template for $b$-jets from the
top-quark decays is fixed assuming it contributes two of the three or
more $b$-tags per observed event.

Background events from non-dilepton $\ttbar$ processes are included
using Monte Carlo simulation, and enter the fit with a fixed
normalization.  Monte Carlo simulation is used to obtain templates for
additional (non-$t \rightarrow Wb$) $b$-jets, $c$-jets, and LF jets.

In the fit to determine the number of $b$-tags from HF jets in
addition to the two $b$-jets from top-quark decay, $N_{\rm HF}$,
separate templates for each category of jet in each of the three
purity classes (high, medium, and low) are used.  The $b$-tagging
efficiencies (Table~\ref{t:tagging_efficiency}) for each flavor of jet
are used to relate the number of jets in each purity bin. After the
application of all constraints, the fit has two floating parameters:
the fraction of LF jets and the fraction of additional $b$-jets. The
fraction of additional $c$-jets makes up the remainder.

Monte Carlo pseudo-experiments show that the fitting method is
unbiased in both best-fit values and estimated uncertainties. The fit
strategy (including estimates of statistical and systematic
uncertainties) was verified using 10\% of the full data sample as well
as with Monte Carlo pseudo-experiments before the fit was performed on
the full data sample. 
These studies indicated that the fit could achieve only a 1$\sigma$
separation of $b$- vs. $c$- jets based on the expected statistical
uncertainty alone. Inclusion of the systematic uncertainty would
further reduce the sensitivity. However, the LF-jet fraction is
expected to be measured with sufficient precision to give a
statistically significant measurement of the total HF content, defined
as the fraction of additional $b$-tagged jets not coming from LF jets.  In
the fit, the individual fractions are not constrained to be positive
or below unity.

\section{Systematic uncertainties}
\label{s:systematics}

Systematic uncertainties may affect the shape of the vertex mass and
$\pt$ templates as well as the acceptance calculations. For the
systematic uncertainties on the template shapes, the fit to the data
is re-evaluated using new templates, derived by varying the relevant
parameters by their systematic uncertainties, and a new fit to the
data is performed. Major uncertainties that affect the fit are the jet
energy scale and resolution, the tagging efficiencies for $b$-, $c$-
and LF jets, the parton-shower and hadronization models, and the Monte
Carlo event generators.

The template for $b$-jets from top-quark decays is nominally taken
from the data with exactly two $b$-tags. To account for kinematic
biases due to additional heavy-flavor jets in the event, a systematic
uncertainty on the shape of this template is assessed using $b$-jets
from top-quark decays from Monte Carlo inclusive $\ttbar$ events with
three or more $b$-tagged jets.

The vertex mass of additional $b$- and $c$-jets is sensitive to the
number of HF quarks contained in a jet (for instance, for $\bbbar$ or
$\ccbar$ produced via gluon splitting). The dominant uncertainty from
this effect would manifest itself as a difference in the shape of the
template for additional $b$-jets. To assess this uncertainty, the
template for additional $b$-jets is replaced by the template for
$b$-jets from top-quark decays.

By default, the normalization of the template for $b$-jets from
top-quark decays is fixed to two per event. A systematic uncertainty
on this normalization is assessed by using the predicted normalization
from Monte Carlo simulation, which includes events with less than two
$b$-tags from top-quark decays, due to $b$-tagging inefficiency. The
total uncertainty due to specific template shape variations is
referred to as `additional fit uncertainties' for the rest of this
paper.

Systematic uncertainties also affect the overall event reconstruction
efficiency. Dominant sources of uncertainty for this category are: the
tagging effiencies for $b$-, $c$- and LF jets, the jet energy scale
and resolution, and the Monte Carlo event generator. Uncertainties on
the lepton identification efficiency, $\met$ reconstruction, and
fragmentation modeling are negligible. In general, systematic
uncertainties are evaluated on the full data sample, with each
uncertainty being taken as the difference between the nominal and the
varied resulting values of $R_{\rm HF}$.

An important uncertainty in this analysis comes from the flavor
composition in the fiducial volume, namely in the value of $F_{b/{\rm HF}}$, 
the fraction of $\ttbar$~+~HF events in the fiducial volume
which contain $b$-jets, used to calculate the correction factor
$\epsilon_{\rm HF}$. 
As described in Sec.~\ref{s:fiducial}, an uncertainty of 10\% on
$F_{b/{\rm HF}}$ is estimated using different Monte Carlo generators. 
It is possible to evaluate $F_{b/{\rm HF}}$ using the data, but with the 
present data set, significant discrimination between $b$- and $c$-jets 
is not possible, making such a comparison of limited use.
Nonetheless, the result of this study is presented as a point of comparison
to the result obtained from the Monte Carlo.

\begin{figure}
\centering
\includegraphics[width=0.51\textwidth]{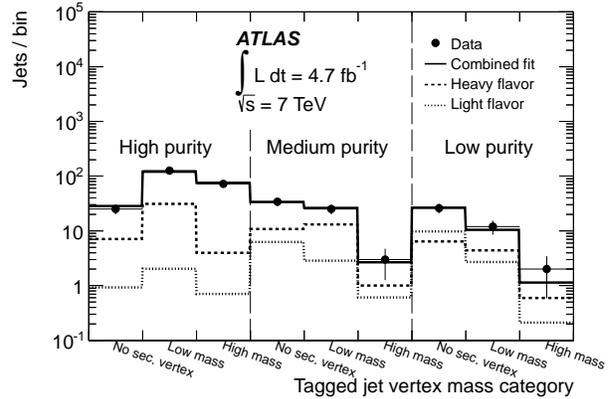}
\caption[Fit template using 4.7 $\invfb$ of data]{The result of the
  template fit (solid line) to the vertex mass distribution in data
  (points). Data are divided into three groups depending on the purity
  of $b$-jets passing each selection, as described in the text. The
  first three bins are the vertex mass distributions for the
  high-purity $b$-tags, the middle three bins for the medium-purity
  $b$-tags, and the last three bins for the low-purity
  $b$-tags. Within each purity category, the first bin contains jets
  with no reconstructed secondary vertex. The middle bin contains jets
  with `low' mass: less than $2~\gev$. The third bin contains jets
  with `high' mass: greater than $2~\gev$. The best fit is shown as a
  sum (labeled as `Combined fit', which includes the $b$-jets from
  top-quark decay) with separate contributions from additional $b$-
  and $c$-jets (labeled as `Heavy flavor'), and LF jets (labeled as
  `Light flavor').}
\label{fig_fitResult_4700}
\end{figure}

\begin{table}[!bhtp]
\centering
\caption[Summary of fit results using 4.7 $\ifb$ of data]{Relative composition
  of $b$-tagged jets in the signal region, fitted in data and compared
  to the expectation from Monte Carlo (MC) simulation. 
In data, the fractions of LF and additional $b$-jets are determined by
the fit. The fraction of $b$-jets from top-quark decays is fixed in
the fit to two $b$-tags in each event. The contributions from $\ttbar$
events with a fake lepton, or non-$\ttbar$ events are fixed in the fit
using the Monte Carlo simulation (those are labeled as `$b$-jets from
other sources' in the table). The fraction of $c$-jets is inferred
from unitarity. All quoted errors are statistical.  }
\begin{tabular}{c|c|c}
\hline
\hline
Type of $b$-tag, fractions & Data fit & MC expectation\\
\hline
Additional LF jets, \%               &   8  $\pm$ 4 & 20   \\
Additional $b$-jets, \%              & $-2$ $\pm$ 7 &  9   \\
Additional $c$-jets, \%              &  26  $\pm$ 8 &  3.5 \\
$b$-jets from $t \rightarrow Wb$, \% &  65          & --   \\
$b$-jets from other sources, \%      &   2.5          & --   \\
\hline
\hline
\end{tabular}
\label{t:fit_result}
\end{table}

\section{Results}
\label{s:results}

In the 106 events in the signal sample (with $\ge 3$ $b$-tagged jets),
there are 325 $b$-tagged jets. After subtracting the non-$\ttbar$
background component, and the contribution from the tagged jets from
the $t \rightarrow Wb$ decay, the number of additional $b$-tags is
found to be 105. 
As described in Sec.~\ref{s:template}, a template fit to all
$b$-tagged jets is performed to determine the flavor composition of
these additional $b$-tagged jets. 
The result of the fit to all 325 $b$-tagged jets is shown in
Fig.~\ref{fig_fitResult_4700}.  The weighted sums of all fit templates
are shown, with contributions for extra HF and mistagged LF jets shown
separately. 
The fitted fractions of $b$-tags from LF jets and additional $b$-jets
are given in Table~\ref{t:fit_result}. Of the 105 additional $b$-tags,
79 $\pm$ 14 (stat.) $\pm$ 22 (syst.) are attributed to HF jets. A
detailed breakdown of the systematic uncertainties on the total number
of HF jets is shown in Table~\ref{t:syst}.

Using Eq.~\ref{eq:xsechf}, the number of HF jets observed in
data, and the quoted correction factor $\epsilon_{\rm HF}$ derived from
the Monte Carlo simulation for $\ttbar$ + HF production, $\sigma_{\rm
  fid}(\ttbar + \rm{HF})$ is found to be 0.16 $\pm$ 0.03 (stat.) pb.
{\sc Alpgen} interfaced with {\sc Herwig} predicts a value of 0.10 pb.

The uncertainty on the fitted fraction of light-flavor jets is
significantly smaller than the uncertainty on the fitted fraction of
additional $b$-jets. This is understood as an effect of fitting in
multiple $b$-purity bins: the low-purity bin is dominated by
light-flavor jets and thus gives improved discrimination. 
The data resolve the total observed HF production rate with a significance
of about $3\sigma$.

\begin{table}[!t]
\caption{Summary of systematic uncertainties (in \%) on the
  measurement of the ratio of fiducial cross sections, $R_{\rm
    HF}$. Uncertainties are quoted separately for the number of HF
  jets measured in the fit ($N_{\rm HF}$), the portion of the
  calculation affecting only the correction factors ($\epsilon_{\rm
    HF}$), and the full calculation. 
As the fit prefers 100\% charm for additional heavy-flavor jets, it
is sensitive to differences in the extra $b$-tagged jets from
the $c$-quark template shape. 
}
\centering
\footnotesize
\begin{tabular}{c|c|c|c}
\hline
\hline
Source                             & \% ($N_{\rm HF}$)& \% ($\epsilon_{\rm HF}$)  & \% (full)             \\
\hline
Lepton reconstruction              &  0.1            & 0.2                          &  0.2                  \\
Jet reconstruction and calibration &  3.5            & 1.6                          &  6.9                  \\
$\met$ reconstruction              &  0.5            & 0.6                          &  0.9                  \\
Fake-lepton estimate               &  3.4            & 0.0                          &  3.4                  \\
\hline%                                                                                                     
Tagging efficiency for $b$-jets    &  1.1            & 2.4                          &  3.1                  \\
Tagging efficiency for $c$-jets    & 25.0            & 5.9                          & 21.2                  \\
Tagging efficiency for light jets  &  8.4            & 0.2                          &  8.4                  \\
\hline%                                                                                                     
Fragmentation modeling             &  6.5            & 15.7                         &  10.2                  \\
Generator variation                &  0.7            & 1.0                          &  1.8                  \\
Initial- and final-state radiation  &  0.1            & 1.7                          &  1.9                  \\
PDF uncertainties                  &  1.6            & 1.0                          &  2.8                  \\
\hline%                                                                                                     
Additional fit uncertainties       &  6.6            & --                           &  6.6                  \\
Fiducial flavor composition        &  0.0            & 6.0                          & 6.0                   \\ 
\hline%                                                                             
Total systematic                   & 29              & 13                           & 28                    \\
\hline
\hline
\end{tabular}
\label{t:syst}
\end{table}

In the data, 1656 $\ttbar$ dilepton candidate events are observed with
at least three jets, at least two of which are $b$-tagged.  The total
background estimate, which is dominated by LF jets misidentified as
$b$-jets from top-quark decay, is found to be 112 $\pm$ 4 (stat.),
leading to a background subtracted yield of 1544 $\pm$ 41 (stat.).
Using Eq.~\ref{eq:xsec}, and the quoted acceptance factor for $\ttbar + j$
production, $\sigma_{\rm fid}(\ttbar + j)$ is found to be 2.55 $\pm$
0.07 (stat.)  pb, compared to 2.83 pb predicted by {\sc Alpgen} and
{\sc Herwig}. 
Taking into account the total uncertainty, it is found that $R_{\rm
  HF}=[6.2 \pm 1.1 ~(\text{stat.}) \pm 1.8 ~(\text{syst.})]$\%. 
A full breakdown of the systematic uncertainties contributing to
$R_{\rm HF}$ is given in Table~\ref{t:syst}.

The extracted value of $\sigma_{\rm fid}(\ttbar$ + HF) is very
sensitive to the value of $F_{b/{\rm HF}}$.  As indicated in
Sec.~\ref{s:fiducial}, the efficiency for $\ttbar + b + X$ events is
approximately a factor of three higher than the corresponding
efficiency for $\ttbar + c + X$ events, implying a potential change in
$\sigma_{\rm fid}(\ttbar$ + HF) by a factor of three if $F_{b/{\rm
    HF}}$ is allowed to vary over the full range [0, 1].

Using the fitted fraction of additional $b$-jets in data results in
$F_{b/{\rm HF}} = -0.02$, with one and two sigma statistical upper
bounds of $F_{b/{\rm HF}} = 0.09$ and 0.27, respectively. This value
is found to be compatible with the predicted value to within 1$\sigma$
when systematic uncertainties are
included. Figure~\ref{fig_acceptance_bcVariation} shows $R_{\rm HF}$
as a function of $F_{b/{\rm HF}}$. The predicted and data-driven
ranges of $F_{b/{\rm HF}}$ are overlaid. With $F_{b/{\rm HF}} = -0.02$
the central value for $R_{\rm HF}$ is determined as 10.7\%.

\begin{figure}
\centering
\includegraphics[width=0.5\textwidth]{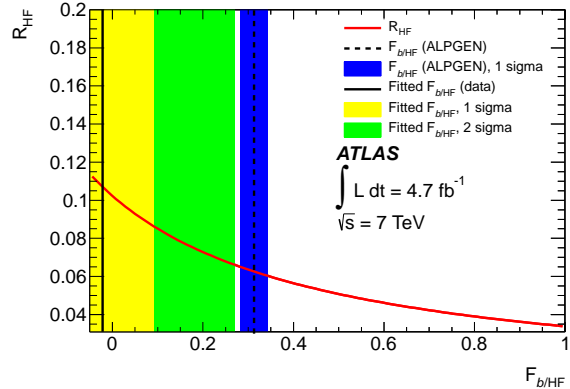}
\caption[Variation of $R_{\rm HF}$ as a function of the $b$:$c$ ratio in
  the fiducial region] {The ratio $R_{\rm HF}$ of fiducial cross
  sections as a function of $F_{b/{\rm HF}}$, the ratio of $\ttbar$ events with additional $b$-quarks
  to $\ttbar$ events with additional $b$- or $c$-quarks. 
  The value of $F_{b/{\rm HF}} = 0.31 \pm 0.03$ is indicated with a
  vertical dashed line with the band around it showing the $1\sigma$
  uncertainty. 
  The fitted fraction of additional $b$-jets is used to extract
  $F_{b/{\rm HF}}$ (solid vertical line) from the data.  The
  statistical uncertainty from the fit is used to define $1\sigma$ and
  $2\sigma$ uncertainty bands.}
\label{fig_acceptance_bcVariation}
\end{figure}

\section{Conclusions}
\label{s:conclusions}

A 4.7 $\invfb$ sample of $7~\TeV$ proton--proton collisions recorded
by the ATLAS detector at the LHC was used to measure the ratio $R_{\rm
  HF}$ of the fiducial cross section for the production of $\ttbar$
events with at least one additional HF quark jet ($\ttbar + b + X$ or
$\ttbar + c + X$) to that for the production of $\ttbar$ events with
at least one additional jet, regardless of flavor, each with $\pt >
25~\gev$ and $|\eta| < 2.5$. A fit to the vertex mass distribution for
$b$-tagged jets in $\ttbar$ candidate events with three or more
$b$-tagged jets is performed to determine the heavy- and light-flavor
content of the additional $b$-tagged jets. The result of the fit shows
that 79 $\pm$ 14 (stat.)  $\pm$ 22 (syst.) of the 105 selected
$b$-tagged jets originate from HF quarks, three standard deviations
away from the hypothesis of zero $\ttbar$~+~HF production. A value of
$R_{\rm HF}=[6.2 \pm 1.1 ~(\text{stat.}) \pm 1.8 ~(\text{syst.})]$\%
is extracted. 
This value of $R_{\rm HF}$ is consistent with the leading order
predictions of 3.4\% obtained from the {\sc Alpgen} Monte Carlo
generator interfaced with {\sc Herwig} and 5.2\% from a calculation
using {\sc Powheg} interfaced with {\sc Herwig}.

\section*{Acknowledgments}

We thank CERN for the very successful operation of the LHC, as well as the
support staff from our institutions without whom ATLAS could not be
operated efficiently.

We acknowledge the support of ANPCyT, Argentina; YerPhI, Armenia; ARC,
Australia; BMWF, Austria; ANAS, Azerbaijan; SSTC, Belarus; CNPq and FAPESP,
Brazil; NSERC, NRC and CFI, Canada; CERN; CONICYT, Chile; CAS, MOST and NSFC,
China; COLCIENCIAS, Colombia; MSMT CR, MPO CR and VSC CR, Czech Republic;
DNRF, DNSRC and Lundbeck Foundation, Denmark; EPLANET and ERC, European Union;
IN2P3-CNRS, CEA-DSM/IRFU, France; GNSF, Georgia; BMBF, DFG, HGF, MPG and AvH
Foundation, Germany; GSRT, Greece; ISF, MINERVA, GIF, DIP and Benoziyo Center,
Israel; INFN, Italy; MEXT and JSPS, Japan; CNRST, Morocco; FOM and NWO,
Netherlands; RCN, Norway; MNiSW, Poland; GRICES and FCT, Portugal; MERYS
(MECTS), Romania; MES of Russia and ROSATOM, Russian Federation; JINR; MSTD,
Serbia; MSSR, Slovakia; ARRS and MVZT, Slovenia; DST/NRF, South Africa;
MICINN, Spain; SRC and Wallenberg Foundation, Sweden; SER, SNSF and Cantons of
Bern and Geneva, Switzerland; NSC, Taiwan; TAEK, Turkey; STFC, the Royal
Society and Leverhulme Trust, United Kingdom; DOE and NSF, United States of
America.

The crucial computing support from all WLCG partners is acknowledged
gratefully, in particular from CERN and the ATLAS Tier-1 facilities at
TRIUMF (Canada), NDGF (Denmark, Norway, Sweden), CC-IN2P3 (France),
KIT/GridKA (Germany), INFN-CNAF (Italy), NL-T1 (Netherlands), PIC (Spain),
ASGC (Taiwan), RAL (UK) and BNL (USA) and in the Tier-2 facilities
worldwide.

\bibliographystyle{myamsplain}
\bibliography{main}

\begin{thebibliography}{10}
\bibitem{atlas_coordinates}
{ATLAS uses a right-handed coordinate system with its origin at the nominal
  interaction point (IP) in the centre of the detector and the $z$-axis along
  the beam pipe. The $x$-axis points from the IP to the centre of the LHC ring,
  and the $y$-axis points upward. Cylindrical coordinates $(r,\phi)$ are used
  in the transverse plane, $\phi$ being the azimuthal angle around the beam
  pipe. The pseudorapidity is defined in terms of the polar angle $\theta$ as
  $\eta=-\ln\tan(\theta/2)$. Transverse momentum and energy are defined as $\pt
  = p\sin\theta$ and $\ET = E\sin\theta$, respectively.}
\bibitem{deltar}
{$\Delta R = \sqrt{(\Delta\phi)^2 + (\Delta\eta)^2}$, where $\Delta\eta$ is the
  separation in $\eta$ between the quark and jet and $\Delta\phi$ is the
  separation in $\phi$.}
\end{thebibliography}

\clearpage
\onecolumngrid
\clearpage 

% ATLAS Collaboration author list
% Data extracted on 12-Nov-2013 for paper reference TOPQ-2012-16
%\documentclass[11pt]{article}
%\usepackage{a4wide}\begin{document}
\begin{flushleft}
{\Large The ATLAS Collaboration}

\bigskip

G.~Aad$^{\rm 48}$,
T.~Abajyan$^{\rm 21}$,
B.~Abbott$^{\rm 111}$,
J.~Abdallah$^{\rm 12}$,
S.~Abdel~Khalek$^{\rm 115}$,
A.A.~Abdelalim$^{\rm 49}$,
O.~Abdinov$^{\rm 11}$,
R.~Aben$^{\rm 105}$,
B.~Abi$^{\rm 112}$,
M.~Abolins$^{\rm 88}$
% ...
\bigskip
\\
% ...
$^{11}$ Institute of Physics, Azerbaijan Academy of Sciences, Baku, Azerbaijan\\
$^{12}$ Institut de F{\'\i}sica d'Altes Energies and Departament de F{\'\i}sica de la Universitat Aut{\`o}noma de Barcelona and ICREA, Barcelona, Spain\\
% ...
$^{21}$ Physikalisches Institut, University of Bonn, Bonn, Germany\\
% ...
$^{48}$ Fakult{\"a}t f{\"u}r Mathematik und Physik, Albert-Ludwigs-Universit{\"a}t, Freiburg, Germany\\
$^{49}$ Section de Physique, Universit{\'e} de Gen{\`e}ve, Geneva, Switzerland\\
% ...
$^{88}$ Department of Physics and Astronomy, Michigan State University, East Lansing MI, United States of America\\
% ...
$^{105}$ Nikhef National Institute for Subatomic Physics and University of Amsterdam, Amsterdam, Netherlands\\
% ...
$^{111}$ Homer L. Dodge Department of Physics and Astronomy, University of Oklahoma, Norman OK, United States of America\\
$^{112}$ Department of Physics, Oklahoma State University, Stillwater OK, United States of America\\
% ...
$^{115}$ LAL, Universit{\'e} Paris-Sud and CNRS/IN2P3, Orsay, France\\
% ...
\end{flushleft}

%\end{document}
% Created with ./xml2latex.py

\end{document}